\documentclass[useAMS,usenatbib]{mn2e}
\usepackage{natbib}
\usepackage{graphicx}
\newcommand{\msun}{\mbox{$\,{\rm M}_\odot$}}

\title[More on the structure of tidal tails]{More on the structure of tidal tails}

\author[A.H.W. K\"upper, R.R. Lane, D.C. Heggie]{Andreas
  H.W. K\"upper$^{1,2}$\thanks{E-mail: \mbox{akuepper@astro.uni-bonn.de} (AHWK); \mbox{rlane@astro-udec.cl (RRL)}; \mbox{dcheggie@ed.ac.uk} (DCH)}, Richard R. Lane$^{3}$ and Douglas C. Heggie$^4$\\
$^{1}$Argelander Institut f\"ur Astronomie (AIfA), Auf dem H\"ugel 71, 53121 Bonn, Germany\\
$^{2}$European Southern Observatory, Alonso de Cordova 3107, Vitacura, Santiago, Chile\\
$^{3}$Departamento de Astronom\'{i}a, Universidad de Concepci\'{o}n, Casilla 160 C, Concepci\'{o}n, Chile\\
$^{4}$University of Edinburgh, School of Mathematics and Maxwell Institute for Mathematical Sciences, 
King's Buildings, Edinburgh EH9 3JZ}

\begin{document}

\date{Accepted \ldots. Received \ldots; in original form \ldots}

\pagerange{\pageref{firstpage}--\pageref{lastpage}} \pubyear{2011}

\maketitle

\label{firstpage}

\maketitle

\begin{abstract}
We investigate the epicyclic motion of stars escaping from star
clusters. Using streaklines, we visualise the path of escaping stars and show
how epicyclic motion leads to over- and underdensities in tidal tails of star
clusters moving on circular and eccentric orbits about a galaxy. Additionally,
we investigate the effect of the cluster mass on the tidal tails, by showing that their structure is better matched when the perturbing effect of the cluster mass is included. By
adjusting streaklines to results of $N$-body computations we can accurately  and quickly
reproduce all observed substructure, especially the streaky features often
found in simulations which may be interpreted in observations as
  multiple tidal tails. Hence, we can rule out tidal shocks as the origin of
  such substructures.  Finally, from the adjusted streakline parameters we
can verify that  for the star clusters we studied escape mainly happens from the tidal radius of the cluster,  given by $x_L = (GM/(\Omega^2-\partial^2\Phi/\partial R^2))^{1/3}$.
We find, however, that there is another limiting radius, the ``edge'' radius, which gives the smallest radius from which a star can escape during one cluster orbit about the galaxy. For eccentric cluster orbits the edge radius shrinks with increasing orbital eccentricity (for fixed apocentric distance) but is always significantly larger than the respective perigalactic tidal radius. In fact, the edge radii of the clusters we investigated, which are extended and tidally filling, agree well with their (fitted) King radii, which may indicate a fundamental connection between these two quantities.
\end{abstract}


\section{Introduction}\label{Sec:Introduction}
Tidal tails of galactic satellites like dwarf spheroidal galaxies and globular
clusters are interesting structures as they tell the story of the dynamical
past of the corresponding object. The most prominent feature of tidal tails is
their large extent across their host galaxy \citep{Yanny03, Grillmair06a,
  Grillmair06b, Grillmair06c}. From the shape of extended tidal tails it is
possible to infer the orbit of their progenitor satellite \citep{Lin77,
  Lynden95, Penarrubia05, Belokurov06, Fellhauer06}. The constraints on the
satellite orbit can be further improved when radial velocity or proper motion
data of tidal tail stars is available \citep{Johnston99, Law05, Binney08,
  Eyre09a, Eyre09b, Odenkirchen09, Koposov10}.

Tidal tails, however, also contain additional information. Variations of the
stellar density along the tails have been observed for extended tidal streams
such as the Milky Way globular cluster Palomar 5
\citep{Odenkirchen03}. For other Milky Way satellites, such as NGC
  288, Willman 1 and NGC 2298, multiple tidal tails have been observed
  \citep{Leon00, Willman06, Balbinot11}. This substructure should also be
somehow linked to the dynamical past of the satellite.

In the common picture, substructure is created by tidal variations, such as
tidal shocks, which temporarily increase the mass loss rate of the
satellite  \citep{Gnedin97}. Tidal tails which are produced in such a violent way should be
dynamically hotter and thus broader than tails originating from dynamical evaporation. Extended satellites like dwarf spheroidal galaxies often show such tidal streams \cite[e.g.][]{Grillmair06d} but tidal tails of star
clusters appear to be typically more compact and colder \citep{Grillmair06a,
  Grillmair06c} suggesting that they are mainly being produced by dynamical
evaporation.

In numerical investigations it has been shown that for star clusters,
tidal variations have to be, in fact, very extreme in order to cause
substructure in their tidal tails \citep{Dehnen04, Kuepper10a}. This is due to
the compactness of star clusters which makes them less vulnerable to tidal
stripping. Observed substructure in tidal tails of  star clusters is,
therefore, most likely not due to tidal variations.

On the contrary, it has been shown that substructure can even arise in tidal
tails of star clusters which move in a steady tidal field \citep{Capuzzo05,
  Kuepper08a, Just09, Kuepper10a}. This substructure is caused by the
epicyclic motion performed by a continuous stream of stars evaporating from
the cluster. It has been shown analytically and numerically that this
substructure can be related to the internal properties of the star cluster
(its mass) and its external properties (its orbit about the galaxy). Moreover,
star clusters on eccentric orbits were also found to produce tidal tail
substructure in this way, rather than through tidal shocks \citep{Kuepper10a}. 

However, clumps in tidal tails may also be produced when parts of a stellar stream pass a spiral arm, a giant molecular cloud or a massive dark-matter subhalo (e.g.~\citealt{Ibata02, Carlberg09}). Therefore, the understanding of the formation of substructure in tidal tails is of vital importance when interpreting observations of clumpy tidal tails.

In the present investigation we show with the help of streaklines exactly how
the epicyclic motion of escaping stars leads to substructure in tidal tails of
star clusters on circular and eccentric orbits (Sec.~\ref{Sec:Theory}). Then
we compare results from $N$-body computations to theoretical streaklines to
visualise the path of escaping stars  and to demonstrate the accuracy of the much more rapid method of streaklines (Sec.~\ref{Sec:Numerical}). In the last
Section we give a brief summary and discuss our results.

\section{Theory}\label{Sec:Theory}
Due to energy equipartition, stars in a star cluster permanently exchange
energy. This process is called two-body relaxation and leads to the continuous
escape of stars from the cluster, since again and again some stars gain
energies above the escape energy (see e.g.~\citealt{Heggie03}). However, in
most cases, even if they are energetically unbound, stars will stay within the
cluster for several dynamical timescales. These ``potential escapers'',
 are preferentially located at large cluster radii where the dynamical timescale is of the order of the orbital
timescale of the star cluster about the Galaxy \citep{Gnedin99, Fukushige00,
  Kuepper10b}. In fact,  a flattening of the velocity dispersion profile has been observed in the
outer parts of Milky Way globular clusters like 47 Tucanae \citep{Drukier98, Scarpa07, Lane09, Scarpa10,
  Lane10a, Lane10b, Scarpa11},  which may well be interpreted as the contribution from potential escapers.

External energy which is added to a cluster, for instance through tidal
shocks, is redistributed within the cluster in the same way. The additional
energy causes more stars to escape from the cluster in a given time interval
but does not immediately influence the structure of the cluster
\citep{Baumgardt03, Kuepper10b}. For this reason tidal shocks do not
inevitably lead to substructure in tidal tails as the escape of unbound stars
is delayed with respect to the tidal shock \citep{Kuepper10a}.  Note that this may be significantly different for extended objects with large relaxation times such as dwarf spheroidal galaxies, especially when they are on very eccentric orbits about their host galaxy (c.f.~\citealt{Penarrubia09}).

Inside the cluster, each star moves within the gravitational field of the
other cluster stars plus the external tidal field of the galaxy. In the
cluster reference frame, in which the cluster is at rest, the motion of stars
is complicated by the Coriolis, centrifugal and Euler pseudo forces resulting
from the cluster's orbital motion about the galactic centre (see
e.g.~\citealt{Chandrasekhar42}). This effective potential of the cluster is
attractive towards the cluster centre within the tidal sphere with radius
$x_L$. This radius gives the distance to the Lagrange points from the cluster
centre along the radial direction connecting the galactic centre with the
cluster centre, i.e. the radius at which the cluster attraction equals the
effective force of the external field. It is often denoted as the tidal radius
and can be approximated  for circular cluster orbits in a Milky Way-like potential as
 \begin{equation}\label{eq:rtc}
x_L \simeq \left(\frac{GM}{2\Omega^2} \right)^{1/3},
 \end{equation}
where $G$ is the gravitational constant, $M$ is the cluster mass, and $\Omega$
is the cluster's angular velocity about the galactic centre  (see e.g.~\citealt{King62, Just09, Kuepper10a}). For eccentric cluster orbits the tidal radius can be calculated using
\begin{equation}\label{eq:rt}
x_L = \left(\frac{GM}{\Omega^2-\partial^2\Phi/\partial R^2} \right)^{1/3},
\end{equation}
where $\Phi$ is the galactic potential and $R$ is the cluster's galactocentric distance (e.g.~\citealt{Kuepper10a}). Due to the dependence on $\Omega$, the tidal radius can change significantly between
peri- and apogalacticon for clusters on eccentric orbits. When a star
evaporates from a cluster it will most likely escape through one of the two
Lagrange points $L_1$ (closer to the galactic centre) or $L_2$ (further away
from the galactic centre), since  escape is easiest through these
points, finally ending up in the leading or trailing tidal tail, respectively.

 Note that the estimate of the limiting radius (Eq.~\ref{eq:rt}) was derived for a star instantaneously situated between the cluster and the galactic centre (see \citealt{King62}). Considerable subsequent research has shown that stars well outside this estimate of the limiting radius may well remain attached to the cluster, for example those on appropriate retrograde orbits \citep{Read06}. Hence, we should keep in mind that at any time there will be stars outside the radius $x_L$ which are actually still bound to the cluster, and that there are stars escaping from radii smaller than $x_L$.

Outside  the tidal radius, the effective potential is repulsive  in some directions such that stars feel
a drag away from the cluster once they escape from its tidal sphere. Hence,
the gravitational field of the cluster quickly becomes unimportant for the
motion of the escapers. Within the tails the stars will perform an epicyclic
motion due to the pseudo forces mentioned above  (plus the tidal force), in which the stars will be
periodically accelerated and decelerated while moving along the tidal
tails. How this epicyclic motion looks depends on the escape conditions of the
stars. In the case where the escape conditions of many escaping stars are
similar, their epicyclic motions will lead to an over- and underdensity
pattern. This is a statistical effect, which is due to the fact that there is
a higher probability for the presence of stars within the epicyclic cusps than
in between two cusps.

In the following we briefly review how the epicyclic motions of escaping stars
depend on the escape conditions. For three different star cluster models we
then show how the orbit influences the epicyclic trajectories of escaping
stars. Moreover, we demonstrate the influence of the cluster mass on these
trajectories. Since the stellar trajectories cannot be calculated analytically
for clusters on eccentric orbits or when the cluster mass is taken into
account, we use streaklines as often applied in fluid dynamics to visualise
the tracks of escaping stars (see Sec.~\ref{sec:streaklines}).

\subsection{Epicyclic motion}\label{sec:epi}
In \citet{Kuepper08a} it was shown for star clusters on circular orbits that,
in a coordinate system in which the cluster is at rest (x-axis points towards
the galactic anticentre, y-axis points into the direction of cluster motion),
the length of an escaper's epicycle along the corresponding tidal tail
(leading or trailing), $y_C$, can be analytically related to the tidal radius,
$x_L$, of the cluster by
\begin{equation}\label{eq:yc}
y_C = \frac{4\pi\Omega}{\kappa}\left(1-\frac{4\Omega^2}{\kappa^2}\right)x_L,
\end{equation}
where $\kappa$ is the epicyclic frequency of the specific galactic
potential. For clusters in Milky Way-like tidal fields $\kappa \simeq
1.4\Omega$ holds \citep{Just09}, such that Eq.~\ref{eq:yc} reduces to
\begin{equation}\label{eq:yc2}
y_C \simeq 3\pi x_L.
\end{equation}
In this calculation it is assumed that the escaping star evaporates from one
of the two Lagrange points and has exactly the same angular velocity,
$\Omega$, as the cluster at the moment of escape. Moreover, the influence of
the gravitational attraction of the cluster is neglected in this ansatz.

\citet{Just09} found a more general solution for the length of the epicycles,
\begin{equation}
y_C = \frac{4\pi}{\beta}\frac{\beta^2-4}{\beta^2}\left(\Delta x+\frac{v_t}{2\Omega}\right),
\end{equation}
where $\beta = \kappa/\Omega$,  $\Delta x$ is some distance from the cluster centre to
the point of escape along the radial direction connecting cluster centre and
galactic centre, and $v_t$ is some additional velocity along the cluster orbit
in the direction of the corresponding tail. For a Milky Way-like potential
this simplifies to
\begin{equation}\label{eq:yc3}
y_C \simeq 3\pi\left(\Delta x+\frac{v_t}{2\Omega}\right),
\end{equation}
which gives Eq.~\ref{eq:yc2} for  $\Delta x=x_L$ and $v_t = 0$. Therefore, if a star
escapes from a radius larger or smaller than the tidal radius, the epicycle
gets larger or smaller, respectively. Furthermore, an additional tangential
velocity in the direction of the corresponding tail will increase the epicycle
length of a star, whereas an additional tangential velocity in the opposite
direction to the corresponding tail will decrease the epicycle length.

The radial offset and the radial amplitude of the epicyclic motion (the
x-component of the epicyclic motion) was derived analytically by
\citet{Just09}. This part of the epicyclic movement depends on the initial
radial offset,  $\Delta x$, and on the additional tangential velocity along the tidal
tails, $v_t$. It is given by
\begin{equation}
x_m = \frac{1}{\beta^2}\left(4\Delta x+\frac{v_t}{\Omega}\pm
\left|\left(4-\beta^2\right)\Delta x + \frac{2v_t}{\Omega}\right| \right),
\end{equation}
where we corrected a typo in the original Just et al.~equation. This amplitude
reduces in a Milky Way-like potential to
\begin{equation}\label{eq:xm}
x_m \simeq 2\Delta x+\frac{v_t}{\Omega}\pm \left| \Delta x+\frac{v_t}{\Omega}\right|,
\end{equation}
that is if $v_t$ is zero then the escapers oscillate around a radius of $2\Delta x$
with an amplitude of $2\Delta x$.

In \citet{Kuepper10a} it was shown with a comprehensive set of $N$-body models
that the epicycles of escaping stars from clusters on circular orbits are on
average a bit larger than predicted when assuming $\Delta x=x_L$ and $v_t = 0$. In
the framework of this theory, this increase in length can be due to both
$\Delta x>x_L$ or $v_t > 0$. In the following we will demonstrate that this
discrepancy is, in fact, due to the neglected cluster mass. Since there is no
simple analytical solution to the problem when the cluster mass is taken into
account, we can only prove this assumption numerically, which will be done
using streaklines.

Furthermore, it was demonstrated in \citet{Kuepper10a} that epicyclic motion
of escaping stars also leads to over- and underdensities in tidal tails of
star clusters moving on eccentric orbits. Even though the shape and density
distribution along the tails gets significantly more complicated for such
clusters, it was shown that the distance of the epicyclic overdensities could
be related to a mean escape radius and a mean escape velocity. Due to the
periodic acceleration and deceleration of the cluster and the tail stars
within the orbit about the galactic centre, the distance from the cluster to
the overdensities periodically increases or decreases, respectively. For this
problem there is no simple analytical solution either, therefore, in the
following we are also going to visualise the trajectories of escaping stars
for such cluster orbits using streaklines.

\subsection{Models}\label{sec:models}
\begin{figure}
\includegraphics[width=84mm]{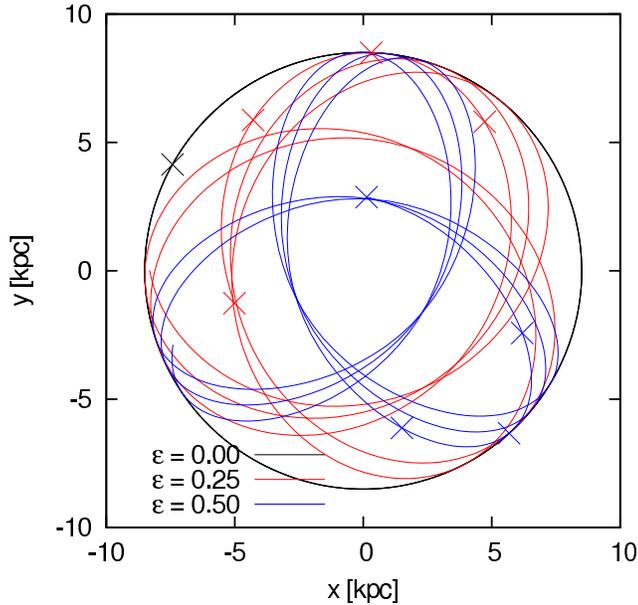}
  \caption{Orbits of the three $N$-body models which were used in this
    analysis for comparison with the theoretical streaklines. One model is on
    a circular orbit, one has an orbital eccentricity of 0.25 and the third
    model follows an orbit with an eccentricity of 0.5. The crosses mark the
    orbital phases where snapshots were taken for analysis (see also
    Tab.~\ref{table1}).}
  \label{orbit}
\end{figure}
\begin{table}
\begin{minipage}{84mm}
\centering
 \caption{Overview of the snapshots from the three cluster models which are
   investigated in detail. The columns give the orbital eccentricity,
   $\epsilon$ (Eq.~\ref{eq:epsilon}), the apogalactic distance, $R_{apo}$, the
   perigalactic distance, $R_{peri}$, and the galactocentric distance,
   $R_{GC}$, of the cluster. The orbital  phase, $p_{orb}$, is defined by
   Eq.~\ref{eq:porb}. The two last columns give the orbital velocity,
   $v_{orb}$, and the angular velocity, $\Omega$, of the cluster at the time
   of the snapshot. }
\label{table1}
\begin{tabular}{ccccccc}
\hline 
$\epsilon$ & $R_{apo}$ & $R_{peri}$ & $R_{GC}$ &  $p_{orb}$ & $v_{orb}$ &$\Omega$\\ 
 &  [kpc] & [kpc] & [kpc] &  & [kms$^{-1}$]& [Myr$^{-1}$]\\ 
\hline 
0.00 & 8.5 & 8.5 & 8.5 & 1.00 & 220 & 0.026\\
\hline
0.25 & 8.5 & 5.1 & 8.5 & 1.00 & 165 & 0.019\\
 &  &  & 7.3 & -0.65 & 205 & 0.028\\
 &  &  & 5.2 & -0.03 & 272 & 0.052\\
 &  &  & 7.5 & 0.71 & 198 & 0.026 \\
\hline
0.50  &  8.5 & 2.8 & 8.5 & 1.00 & 110 & 0.013\\
  &   &  & 6.6 & -0.67 & 189 & 0.029\\
  &   &  & 2.9 & 0.02 & 330 & 0.114\\
  &   &  & 6.3 & 0.61 & 201 & 0.032\\
\end{tabular}
\end{minipage}
\end{table}
We investigate the epicyclic motion of tail stars for three different types of
cluster orbits. First of all, we look at the circular orbit case, then we look
at two eccentric orbits with eccentricities, $\epsilon$, of 0.25 and 0.5. The
eccentricity is defined as
\begin{equation}\label{eq:epsilon}
\epsilon = \frac{R_{apo}-R_{peri}}{R_{apo}+R_{peri}},
\end{equation}
where $R_{apo}$ is the cluster's apogalactic distance and $R_{peri}$ its
perigalactic distance. A circular orbit therefore has an eccentricity of 0.

For the integration of the orbits we use the Milky Way-like potential
suggested by \citet{Allen91}. This galactic potential, $\Phi$, consists of a central
point-mass potential given by
\begin{equation}
\Phi_c (R) = -\frac{M_1}{\sqrt{R^2+b_1^2}},
\end{equation}
a \citet{Miyamoto75} disk potential given by
\begin{equation}
\Phi_d (x,y,z) = -\frac{M_2}{\sqrt{x^2+y^2+\left(a_2 + \sqrt{z^2+b_2^2}\right)^2}},
\end{equation}
and a halo potential of the form
\begin{eqnarray}
\Phi_h (R) &=& -\frac{M(R)}{R}-\frac{M_3}{1.02a_3}\\
&&\left[-\frac{1.02}{1+(R/a_3)^{1.02}}+\ln\left(1+(R/a_3)^{1.02}\right)\right]_R^{100},\nonumber
\end{eqnarray}
where
\begin{equation}
M(R) = \frac{M_3(R/a_3)^{2.02}}{1+(R/a_3)^{1.02}}.
\end{equation}
The numerical constants $M_1$, $b_1$, $M_2$, $a_2$, $b_2$, $M_3$ and $a_3$ are chosen such that the combined potential of the three components has a circular velocity of 220
kms$^{-1}$ at 8.5 kpc (see \citealt{Allen91}). 

The apogalactic distance of all three clusters is 8.5
kpc (which is a somewhat arbitrary choice). Therefore, the circular orbit has
a constant orbital velocity, $v_{orb}$, of 220 kms$^{-1}$, whereas the
eccentric orbits have an orbital velocity of 165 kms$^{-1}$ and 110 kms$^{-1}$
at apogalacticon for an eccentricity of 0.25 and 0.5, respectively. Their
perigalactic distances are 5.1 kpc and 2.8 kpc, respectively, and their
velocities at these orbital phases are 275 kms$^{-1}$ and 334 kms$^{-1}$. The
orbital timescale is 240 Myr for the circular case, 133 Myr for $\epsilon =
0.25$ and 113 Myr for $\epsilon = 0.5$. An overview of these quantities is
given in Tab.~\ref{table1}. In Fig.~\ref{orbit} the integrated orbits are
shown for 400 Myr.

The shape and structure of tidal tails for star clusters on eccentric orbits
is strongly influenced by the orbital phase of the cluster \citep{Kuepper10a,
  Kuepper11a}. To illustrate the changes between the different orbital phases,
we pick four snapshots from the integrations of each of the two eccentric
orbits. The snapshots are taken in steps of 30 Myr starting from
apogalacticon. In the next snapshots the clusters are in an orbital phase
between apogalacticon and perigalacticon in which the clusters are accelerated
towards the galactic centre. The third snapshots are close to perigalacticon,
and the last ones are between perigalacticon and apogalacticon in which the
clusters and their tails are decelerated. The points at which those snapshots
were taken are illustrated in Fig.~\ref{orbit} as crosses.

One way to quantify the orbital phase, $p_{orb}$, of a star cluster in its
orbit about the galaxy is given in \citet{Kuepper11a} by
\begin{equation}\label{eq:porb}
p_{orb} = \frac{\dot{R}_{GC}}{|\dot{R}_{GC}|}\frac{R_{GC}-R_{peri}}{R_{apo}-R_{peri}},
\end{equation}
where $R_{GC}$ is the cluster's current galactocentric radius, and
$\dot{R}_{GC}$ is the time derivative of this radius. In this way, the orbital
phase is zero at perigalacticon and unity at apogalacticon  (in fact $\pm 1$). Moreover, it is
negative when the cluster is moving to perigalacticon and positive if its
moving to apogalacticon. The orbital phases of the cluster snapshots which we
investigate in detail are given in Tab.~\ref{table1}.

\subsection{Streaklines}\label{sec:streaklines}
\begin{figure*}
\includegraphics[width=84mm]{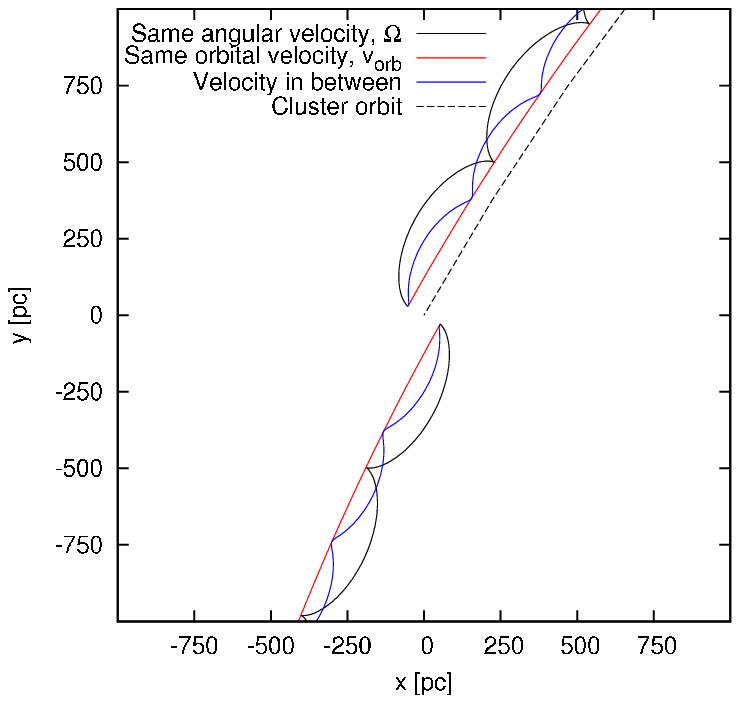}
\includegraphics[width=84mm]{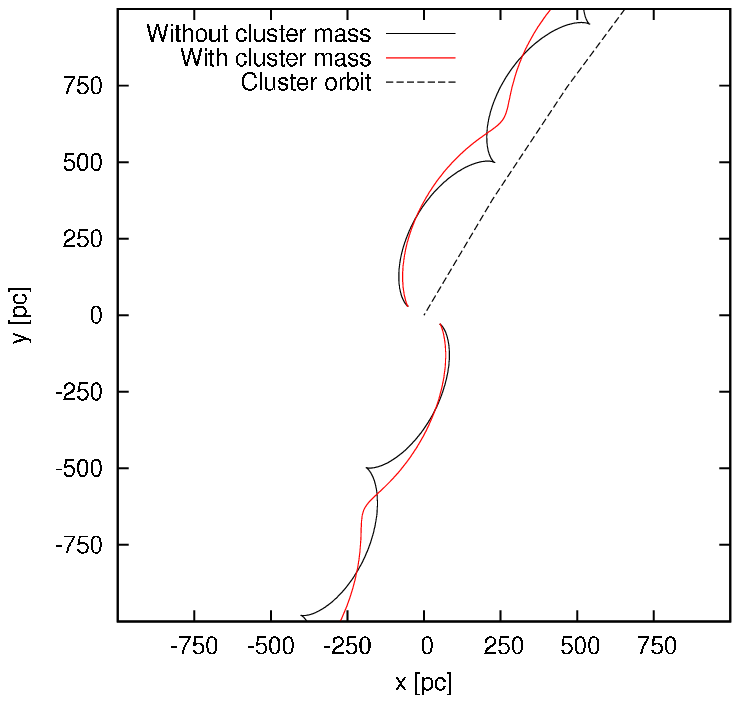}
  \caption{Snapshot of the cluster model on a circular orbit (black dashed
    lines). Left panel: to produce the three sets of streaklines, test
    particles were released during the last few hundred Myr of this numerical
    integration from two radial offsets, $\pm \Delta x$  (which we chose to be 60 pc in this example), with respect to the cluster
    orbit, and with three different tangential velocities, $v_t$. The black
    solid lines show the streaklines resulting from the test particles being
    released with the same angular velocity as the cluster. The epicyclic
    motion as described in \citet{Kuepper08a} is nicely reproduced. The red
    solid lines show the path of the test particles which have the same
    orbital velocity as the cluster. Just like the cluster they move on
    circular orbits about the galactic centre. The blue solid lines show the
    test particles with velocities in between these two cases. The length of
    the epicycles is shorter and the radial amplitude is smaller. The cluster
    mass was neglected in these calculations. Right panel: here the effect of
    the cluster mass on the trajectories of the test particles is shown.  The cluster mass was chosen such that the tidal radius is 60 pc. The black solid lines are the same as in the left panel, whereas the red solid
    lines show the influence of the cluster mass: the sharp cusp is replaced
    by a smooth minimum, which, in addition, is further away from the
    cluster. Moreover, the test particles are, on average, further away from
    the cluster orbit in radial direction.}
  \label{streak_circular}
\end{figure*}
\begin{figure*}
\includegraphics[width=168mm]{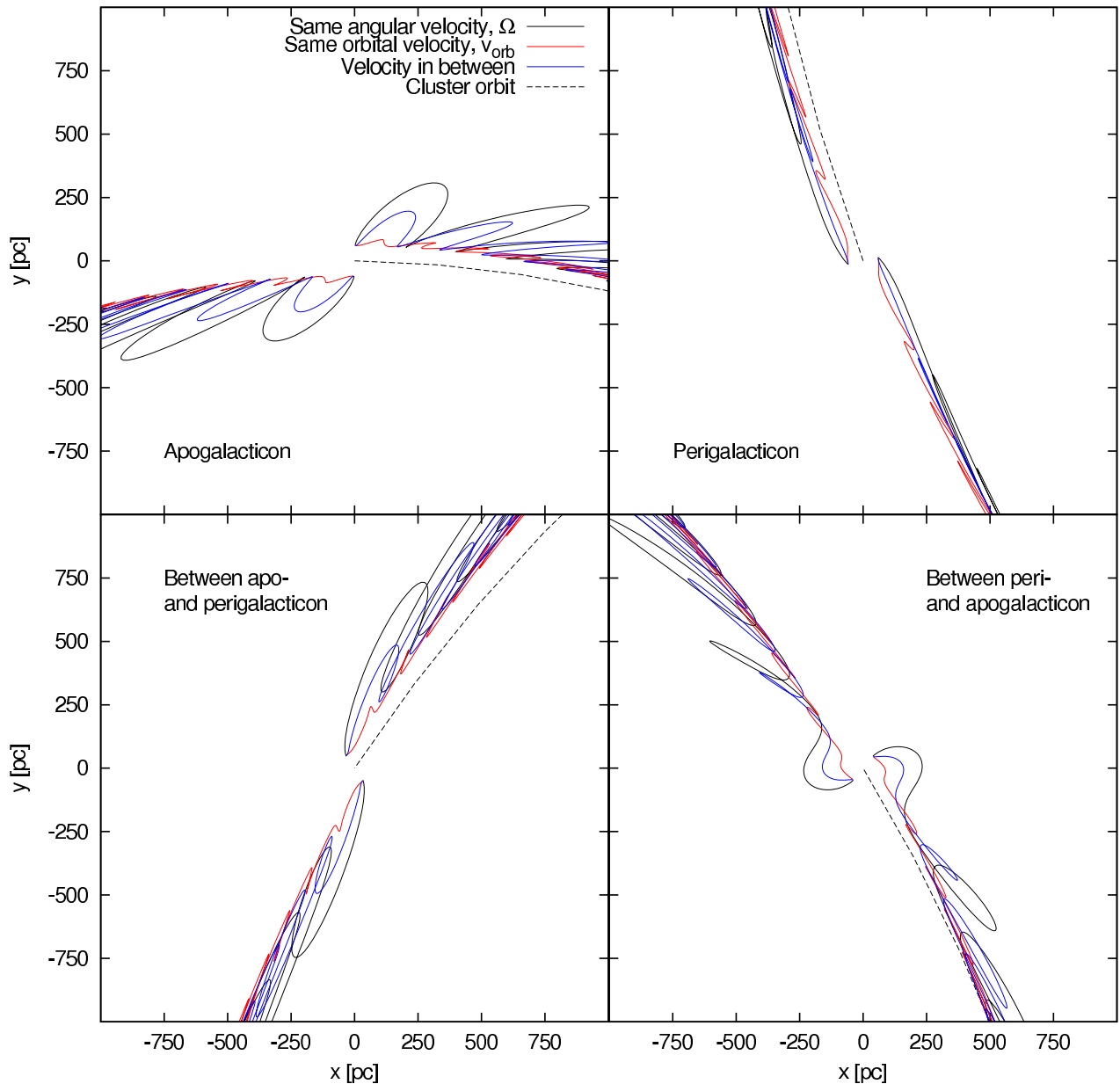}
  \caption{The same as in the left panel of Fig.~\ref{streak_circular} but for
    the cluster with an orbital eccentricity of 0.25. The four snapshots show
    the cluster and the corresponding streaklines in four different orbital
    phases (see also Tab.~\ref{table1}).  $\Delta x$ was chosen to be 60 pc just like in Fig.~\ref{streak_circular}. The epicyclic movement gets distorted
    due to the eccentric cluster orbit. At apogalacticon (upper left panel)
    the epicycles are squeezed together, whereas at perigalacticon (upper
    right panel) the streaklines are stretched. In between these two orbital
    phases the shape of the streaklines differs if the cluster and its tails
    are being accelerated (lower left panel) or decelerated (lower right
    panel). Moreover, we get epicyclic movement also for the case when the
    test particles are released from the cluster with the same orbital
    velocity as the cluster (red solid lines). }
  \label{1204_streak}
\end{figure*}
\begin{figure*}
\includegraphics[width=168mm]{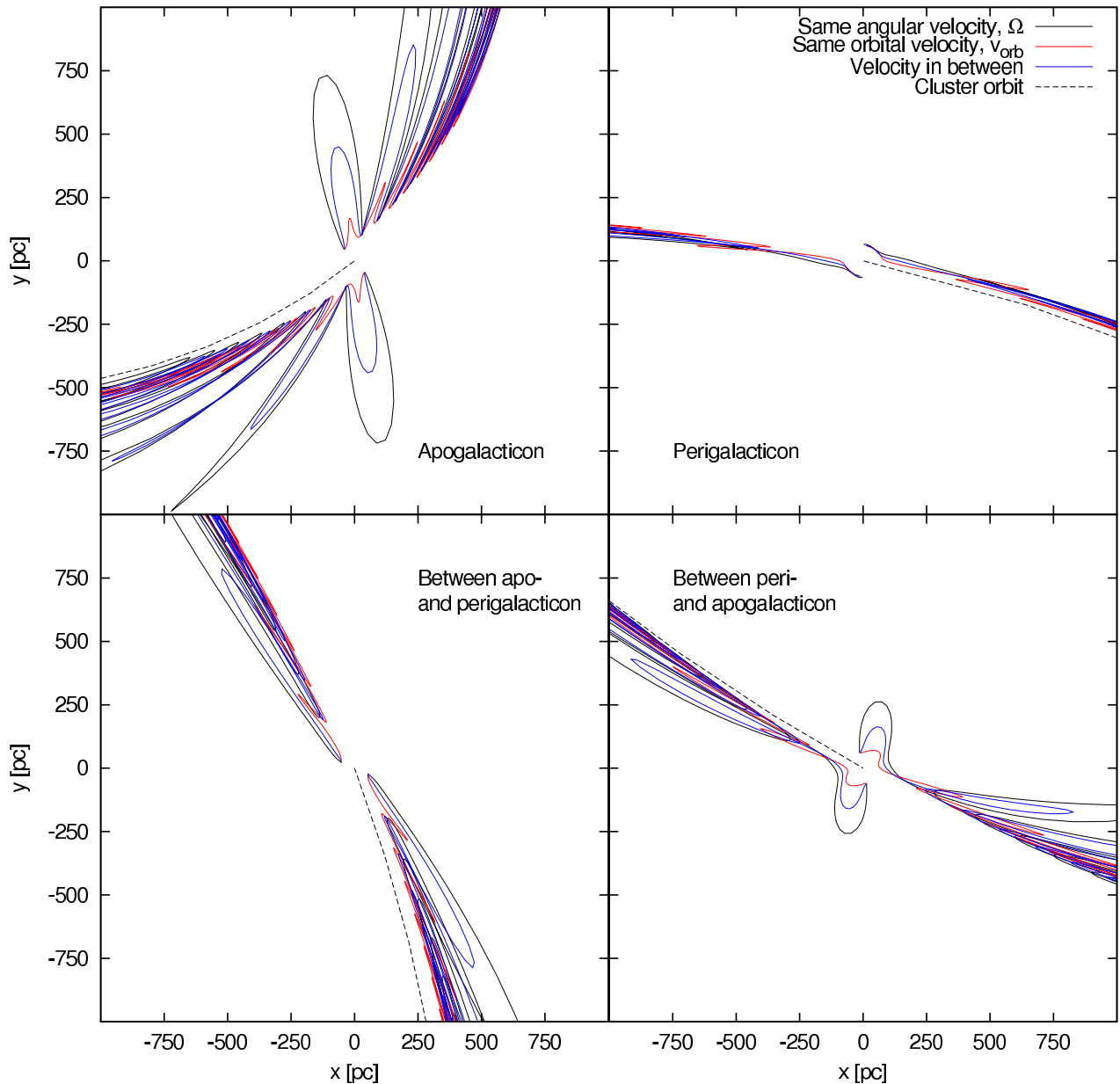}
  \caption{The same as Fig.~\ref{1204_streak} but for an orbital eccentricity
    of 0.5. Again, the four snapshots show four different orbital phases (see
    also Tab.~\ref{table1}).   In this case $\Delta x$ was also chosen to be 60 pc just like in Fig.~\ref{streak_circular}. We see that the distortion effects get stronger
    with increasing eccentricity. The compression at apogalacticon is stronger
    than in Fig.~\ref{1204_streak}, and so is the stretching at
    perigalacticon. Moreover, we can observe the effect of differential
    acceleration along the cluster-tail system. For example in the lower right
    panel the leading tail (to the right) is broader than the trailing tail.}
  \label{1203_streak}
\end{figure*}
\begin{figure*}
\includegraphics[width=168mm]{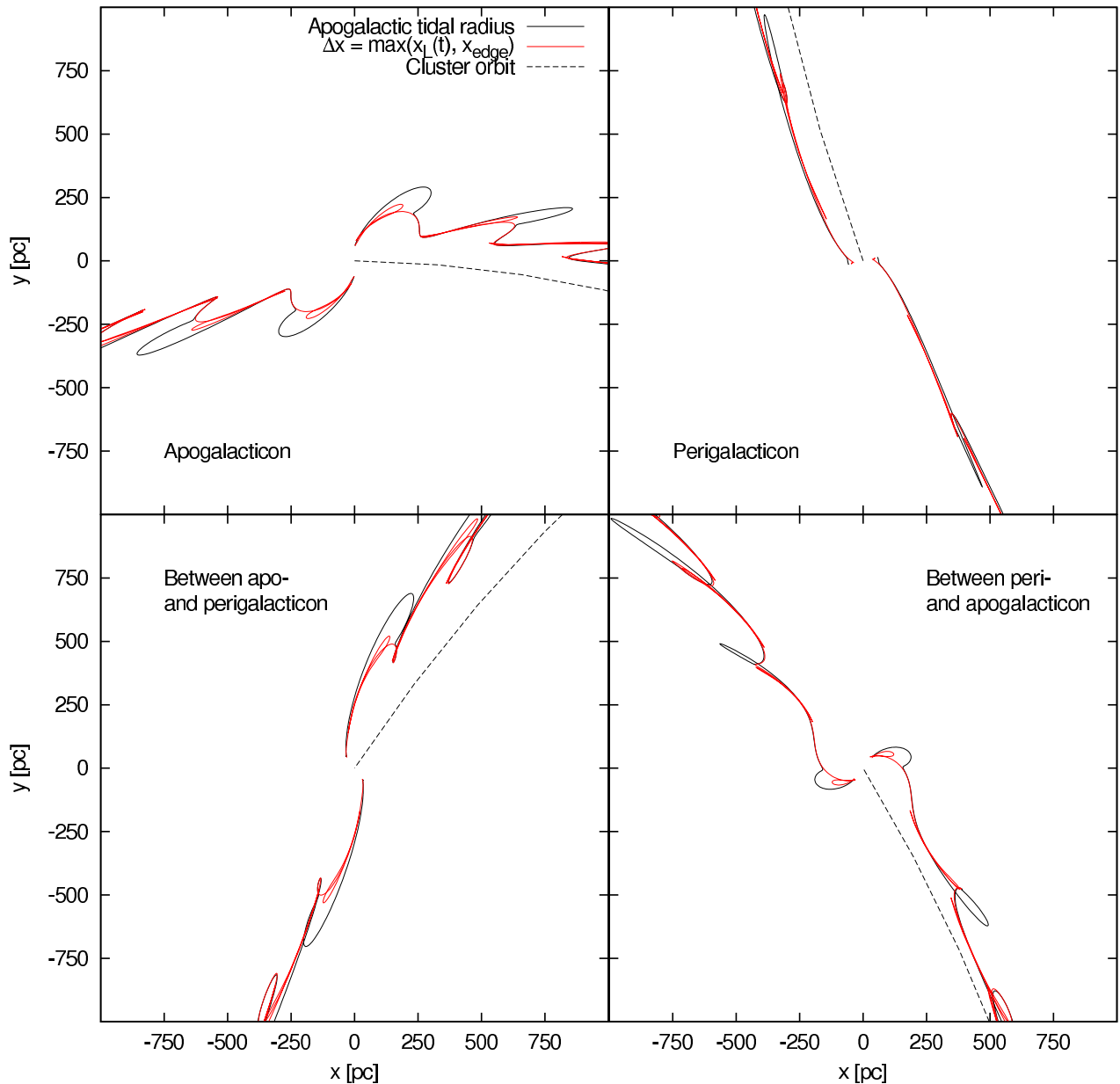}
  \caption{The same as Fig.~\ref{1204_streak} but here the cluster mass is
    taken into account. For the black solid lines the test particles were
    released from the cluster's apogalactic tidal radius.   The cluster mass was chosen such that the apogalactic tidal radius is 60 pc. The lines are
    similar to the lines for which the cluster mass was neglected (black solid
    lines in Fig.~\ref{1204_streak}). When the test particles are released
    from the actual tidal radius of the cluster (red solid lines), the
    influence of the cluster mass gets more pronounced, as can be seen in the
    additional loops and bends. To produce these streaklines we had to
    introduce a minimum radius  from which particles were released, $x_{edge}$, since else some test particles would
    have been re-captured by the cluster at a later orbital phase. This
    ``edge'' radius is  49.0 pc, whereas its
    perigalactic tidal radius is 37.5 pc.}
  \label{1204_streak2}
\end{figure*}
\begin{figure*}
\includegraphics[width=168mm]{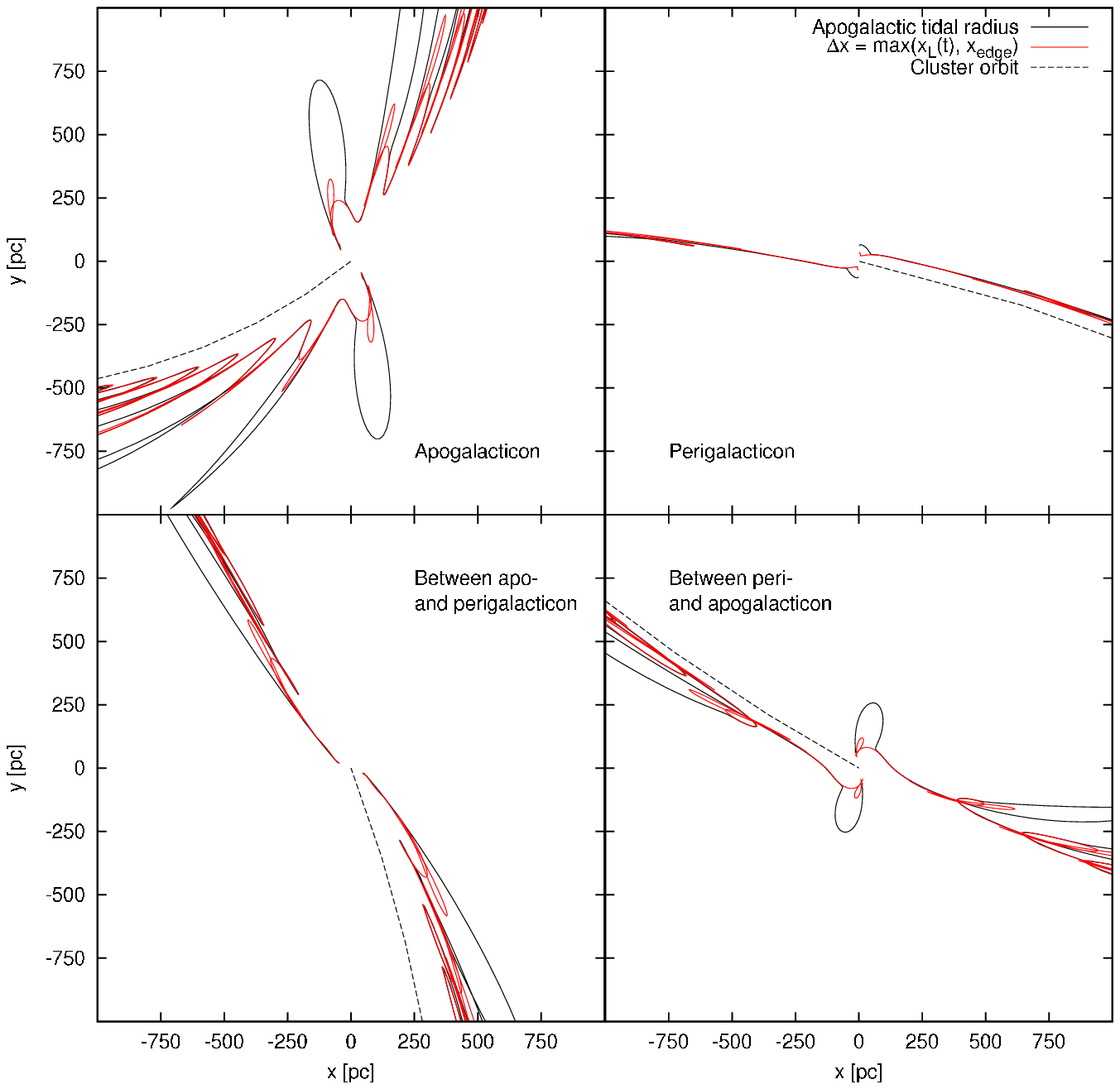}
  \caption{The same as Fig.~\ref{1204_streak2} but for the cluster on an orbit
    with eccentricity of 0.5.  Again, the cluster mass was chosen such that the apogalactic tidal radius is 60 pc. The streaklines for which the test particles
    were released from  the apogalactic tidal radius (black solid lines) are again similar to
    the streaklines for which the cluster mass was neglected (black solid
    lines in Fig.~\ref{1203_streak}). When the test particles are released
    from the actual tidal radius the streaklines get much more complex (red
    solid lines). For these streaklines we had to assume an edge radius 32 pc, whereas the perigalactic tidal
    radius is 21.2 pc.}
  \label{1203_streak2}
\end{figure*}
In \citet{Kuepper08a} and \citet{Just09} it was shown that the orbits of stars
escaping from a cluster can, under certain assumptions, be integrated
analytically. In this simplified case a star is released from the cluster
centre with a certain positive or negative offset along the galactocentric
radius, and with some negative or positive offset velocity with respect to the
cluster. One of the main assumptions here is that the cluster moves in a
circular orbit about the galactic centre. Moreover, the cluster mass is
neglected in this calculation.

For eccentric cluster orbits such a general analytical solution does not
exist. In addition, the trajectories of escaping stars depend on the orbital
phase of the cluster in this case. Instead of solving the equations of motion
analytically, we can simply illustrate the trajectories of stars for eccentric
cluster orbits by using streaklines. This is a concept from fluid dynamics in
which test particles are released into a fluid from a given point to visualise
the flow. In engineering streaklines are often produced with smoke or dye.

To produce the streaklines we integrate the cluster orbit and, at given time
intervals, release test particles. These particles have a certain velocity,
$v$, and are released from two points which are given by the cluster centre
plus a fixed positive or negative radial offset, $\Delta x$, along the galactocentric
radius vector. All released test particles are then integrated together with
the cluster to the point in time of the snapshot. The gravitational influence
of the cluster on the test particles is taken into account in the cases where
it is mentioned explicitly. There is no interaction between the test
particles.

For this purpose, it is not reasonable to stay within the rotating,
accelerated reference frame in which a cluster on a circular orbit is at rest,
for which the analytical solutions of \citet{Kuepper08a} and \citet{Just09}
have been developed. Thus, we have to keep in mind that we have to convert the
tangential offset velocities in the rotating coordinate system, $v_t$, to the
Cartesian, galactocentric rest frame velocities, $v$, when applying
Eq.~\ref{eq:yc3}.

First we illustrate that the technique works correctly by comparing it with
results from the formalism described in Sec.~\ref{sec:epi} for circular
orbits. Therefore, we produce three sets of streaklines in which the test
particles are released with three different velocities, $v$, from the cluster
on a circular orbit:
\begin{enumerate}
\item\label{v1} the test particles are released with the same angular velocity as the
  cluster, $\Omega$, i.e.~the test particles which are released from the
  larger galactocentric radius have a higher orbital velocity than the
  cluster, $v = v_{orb} + \Omega \Delta x$, whereas the test particles at the smaller
  galactocentric radius have a lower orbital velocity, $v = v_{orb} - \Omega
  \Delta x$. This corresponds to the case $v_t = 0$ in Eq.~\ref{eq:yc3},
\item\label{v2} the test particles have the same orbital velocity as the cluster, i.e. $v = v_{orb}$. This corresponds to $v_t = -\Omega \Delta x$ in Eq.~\ref{eq:yc3},
\item\label{v3} the test particles have an orbital velocity in between the two cases,
  i.e. $v_t =- 0.5\Omega \Delta x$ corresponding in the galactocentric rest frame to
  $v = v_{orb} +0.5\Omega \Delta x$ for the leading tail and $v = v_{orb} - 0.5\Omega
  \Delta x$ for the trailing tail, respectively.
\end{enumerate}
These sets of streaklines we also produce for the eccentric orbits to see how
the orbital acceleration and deceleration influence the trajectories of
escaping stars.

Finally, we produce additional sets of streaklines for which the mass of the
cluster is taken into account. The cluster mass has so far been neglected but
will, of course, influence the trajectories of escaping stars. The cluster can
attract stars and even re-capture them if they happen to lie beyond the tidal
radius (given by Eq.~\ref{eq:rt}) at some later orbital phase. \citet{Read06}, for example, demonstrated that stars orbiting the cluster on prograde orbits with respect to the cluster motion about the galactic centre can escape the cluster potential more easily than stars on retrograde orbits. We therefore have to consider from which radius we release the test particles to produce the streaklines.

If we chose to release the test particles from the apogalactic tidal radius,
which is the largest tidal radius throughout the orbit since the cluster has
the lowest angular velocity at this point (see Tab.~\ref{table1}), then we  may expect that virtually all test particles will escape the gravitational attraction of
the cluster mass. This, however, might not be the most realistic approach. If
we, on the other hand, chose to release the test particles always  from the actual tidal radius, $x_L(t)$ (Eq.~\ref{eq:rt}), at the given time, $t$, then many test particles may be
re-captured. This will especially be a problem for the test particles released
near perigalacticon since the tidal radius is smallest here and, in addition,
changes rapidly in this orbital phase as the angular velocity is largest at
this point.

As we will demonstrate later, for this reason it is necessary to introduce a
minimum radius from which test particles can be released such that no test
particles gets re-captured. This radius we call the ``edge'' radius, $x_{edge}$, and it
serves as a lower limit to the actual radius from which test particles will be released, i.e.
\begin{equation}\label{eq:xoft}
\Delta x (t) = \max\left( x_L(t), x_{edge}\right).
\end{equation}
We determine the value of this edge radius by
incrementally increasing the lower limit from  the value of the
perigalactic tidal radius until no test particle is re-captured.

In the following we will show that, for circular cluster orbits, this edge
radius coincides with the constant tidal radius, since any particle released  from rest
from a smaller cluster radius will not be able to escape from the cluster
potential  (for the given escape conditions \ref{v1}-\ref{v3}). For eccentric orbits this value is smaller than the apogalactic
tidal radius but significantly larger than the perigalactic tidal radius. Its
value decreases with increasing eccentricity  for fixed apocentric distance.

\subsection{Circular orbit}\label{sec:circular}
In the left panel of Fig.~\ref{streak_circular} the resulting streaklines for
the circular cluster orbit are shown. The dashed line gives the orbit of the
cluster. Along this orbit test particles have been released in constant time
intervals with a positive or negative spatial offset, $\Delta x$, from the cluster
centre along the galactocentric radius vector and with one of the three
velocity offsets, $v_t$, given above.  We chose $\Delta x$ to be 60 pc, which is a somewhat arbitrary choice. The cluster and the test particles have each been independently integrated to the point in time of the snapshot.
Hence, there is no gravitational influence of the cluster on the test
particles.

From the left panel of Fig.~\ref{streak_circular} it can be seen that the
analytical results of \citet{Kuepper08a} and \citet{Just09} are nicely
reproduced when the test particles are released with the same angular velocity
as the cluster (black solid line). Moreover, if the orbital velocity of the
test particles at the larger (smaller) galactocentric radius is lower (higher)
than in the case of equal angular velocity, the length of the epicycle is
shorter (blue solid line).

In the case when the test particles have the same orbital velocity as the
cluster then the epicycles should, in principle, be half as short as in the
case of sharing the cluster's angular velocity, since
\begin{equation}
y_C = 3\pi\left( \Delta x+\frac{-\Omega\Delta x}{2\Omega}\right) = 3\pi \frac{\Delta x}{2}.
\end{equation}
Instead, we see that the the test particles follow a circular orbit (red solid
line). From Eq.~\ref{eq:xm} we see that, in the case when $v_t = -\Omega \Delta x$,
the radial  extrema in the epicycles are at
\begin{equation}
x_m = 2\Delta x+\frac{v_t}{\Omega}\pm \left| \Delta x+\frac{v_t}{\Omega}\right| = \Delta x\pm 0.
\end{equation}
Hence, the radial amplitude is zero, and the test particles are moving on a
circular orbit about the galactic centre with a radial offset to the cluster
of size $\Delta x$.

The theoretical framework (Eq.~\ref{eq:yc}--\ref{eq:xm}) is in perfect
agreement with our streaklines. Now we include the gravitational attraction of
the cluster in our computations. The mass of the cluster in this computation
is chosen such that the tidal radius, $x_L$ (Eq.~\ref{eq:rt}), coincides with
the starting point of the test particles, $\Delta x$,  i.e.~60 pc in our example.

As we can see in the right panel of Fig.~\ref{streak_circular}, the cluster
mass increases the length of the epicycles, $y_C$, by about 25\%. Moreover, the sharp cusp at
$y_C$ is replaced by a smooth minimum, and the radial distance of the test
particles from the cluster orbit, $\Delta x$, is larger on average. The radial amplitude appears to be damped by a factor of about 0.6 when the cluster mass is taken into account.

All these effects can be explained when looking at the shape of the effective
potential of cluster and galaxy near the tidal radius (see e.g.~figure 2 of
\citealt{Just09}). Beyond the tidal radius the effective potential is
repulsive such that the test particles get accelerated away from the
cluster. \citet{Just09} found that, due to Jacobi-energy conservation (see
e.g.~\citealt{Heggie03}), a test particle starting at $\Delta x=x_L$ with $v_t = 0$
will be at $\Delta x=\sqrt{3}x_L$ if it ever comes to rest again in the future. In
fact, the test particles in the minimum of our example are less than 10\% off
this value as they still have a velocity of 0.5 km/s with respect to their
initial conditions. 

Note that, since the distance of the Lagrange points as well as the length of the epicycles scales with the mass of the cluster, the effect of the cluster mass on the trajectories will always be the same. Stars escaping from a larger radius than the tidal radius will, of course, be less influenced by the cluster mass. Their trajectories will look more like the ones for which we neglected the cluster mass.

\subsection{Eccentric orbits}\label{sec:eccentric}
In Fig.~\ref{1204_streak} the streaklines of the cluster in an eccentric orbit
with eccentricity 0.25 are shown for four different orbital phases. Here,
again, the cluster mass was neglected.  Also, we chose $\Delta x$ to be 60 pc like in the previous example. We see that the streaklines are distorted in comparison to the circular orbit case
(Fig.~\ref{streak_circular}). Moreover, the distortion depends on the orbital
phase of the cluster. At apogalacticon the streaklines appear to be similar to
the circular orbit case but squeezed together. The length of the epicycles is
much shorter than in the circular orbit case and epicyclic loops further away
from the cluster are tilted towards the cluster orbit.

Between apo- and perigalacticon, when the cluster and the test particles are
accelerated towards the galactic centre, the tilt of the epicyclic loops gets
stronger, whereas the length of the epicycles increases. At perigalacticon the
epicyclic loops are almost completely aligned with the cluster orbit. The
length of the epicycles is largest in this orbital phase. Between peri- and
apogalacticon, when the cluster and the test particles are decelerated, the
tilt gets weaker, the epicycles shorter and the loop structure gets more
pronounced.

Also apparent from Fig.~\ref{1204_streak} is the thickness of the tidal
tails. Assuming that most stars follow such an epicyclic motion, we can expect
the tidal tails to be thickest at apogalacticon and thinnest at
perigalacticon. Moreover, we see that we get epicyclic motion even when the
escaping stars have the same orbital velocity as the cluster. This is due to
the stars and the cluster being on slightly different eccentric orbits about
the galactic centre. The same is shown in Fig.~\ref{1203_streak} for the
cluster with an orbital eccentricity of 0.5. All effects described above get
more pronounced. Furthermore, with increasing orbital eccentricity it gets
more difficult to separate the individual streaklines as the epicyclic loops
overlap multiply.

In these illustrations the offset radius, $\Delta x$, from which the test particles
have been released were the same for all snapshots. When we take the cluster
mass into account in our computations then we can fix $\Delta x$ to the apogalactic
tidal radius in order to be able to better compare the resulting streaklines
to Fig.~\ref{1204_streak} \& \ref{1203_streak}.  Therefore we chose the cluster mass such that the apogalactic tidal radius is 60 pc. The streaklines are shown in
Fig.~\ref{1204_streak2} (black solid lines) for the cluster on an orbit with
eccentricity of 0.25. As we can see, the streaklines look quite similar to the
lines for which the cluster mass was neglected (black solid lines in Fig.~\ref{1204_streak}). Only some minor
bends in the lines tell us that there is some influence of the cluster on the
test particles.

This changes when we release particles from the actual tidal radius
(Eq.~\ref{eq:rt}). The influence of the cluster mass on the test particles
gets much more obvious, as the lines show additional loops and bends (red
solid lines). To produce the streaklines in Fig.~\ref{1204_streak2} we had to
introduce a minimum radius  from which we release particles, the ``edge'' radius, $x_{edge}$, mentioned above
(Sec.~\ref{sec:streaklines}, Eq.~\ref{eq:xoft}). For the orbit with an eccentricity of 0.25 this
edge radius  was found to be 49.0 pc (about 80\% of the apogalactic tidal radius), whereas its perigalactic
tidal radius is 37.5 pc.

Both sets of streaklines are shown in Fig.~\ref{1203_streak2} for the cluster
on an orbit with an eccentricity of 0.5.  Here again we adjusted the cluster mass such that the apogalactic tidal radius is 60 pc. The effects described above get more
pronounced. For the black solid lines the test particles were released from
the apogalactic tidal radius, and for the red solid lines they were released
from the actual tidal radius. The edge radius, which we had to introduce here
as well, is  32.0 pc (about 50\% of the apogalactic tidal radius), whereas the perigalactic
tidal radius is 21.2 pc.

\section{Comparison with $N$-body data}\label{Sec:Numerical}
\begin{figure}
\includegraphics[width=84mm]{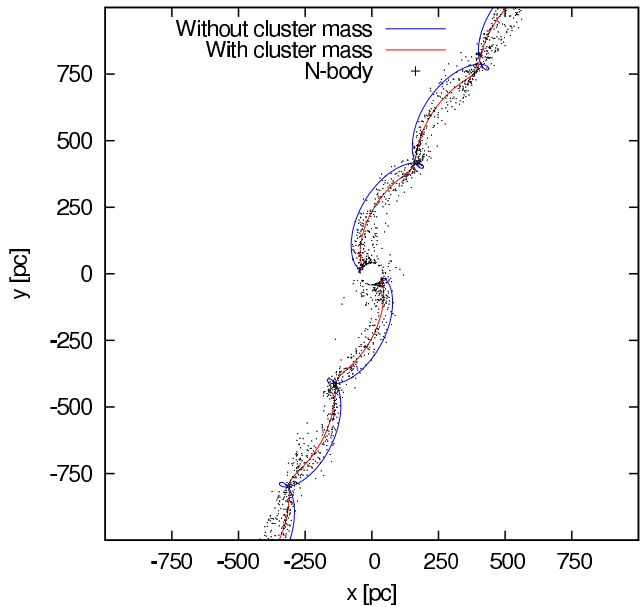}
  \caption{Comparison of streaklines to results from $N$-body computations of
    a cluster on a circular orbit. The cluster was integrated for 2 Gyr and
    has lost about $3600\msun$ of stars which are mainly in the tidal tails
    now (black dots; stars within a cluster radius of 40 pc were omitted for
    clarity). The blue solid lines show the streaklines for which the cluster
    mass was neglected. To match the length of the epicyclic overdensities,
    the test particles were released with a tangential velocity of $v_t =
    0.55$ kms$^{-1}$ from the cluster's tidal radius, i.e.~$\Delta x=x_L=37$ pc. For
    the red solid lines the cluster mass was taken into account. The test
    particles were released from the cluster's tidal radius with zero
    velocity. As we can see, the shape of the tails is much better reproduced
    by the streaklines which take the cluster mass into account. The scatter
    of stars about this streakline may well originate from the scatter in
    escape conditions.}
  \label{1202_sim_circular}
\end{figure}
\begin{figure}
\includegraphics[width=84mm]{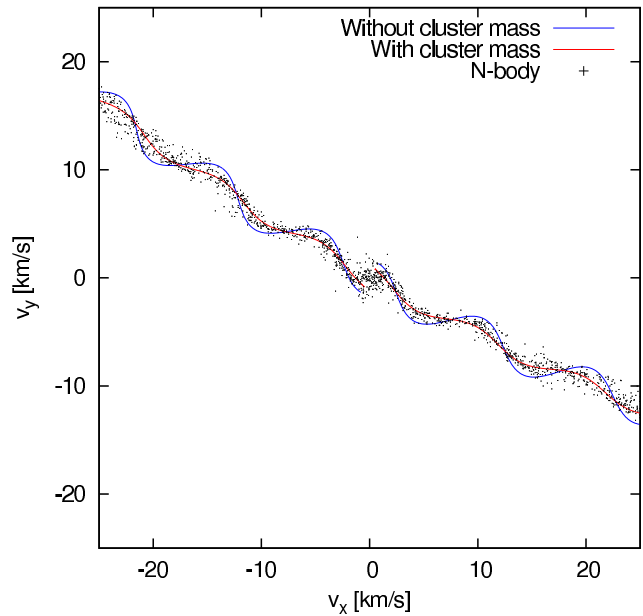}
  \caption{The same as Fig.~\ref{1202_sim_circular} but in velocity space (velocities are given with respect to the cluster velocity). Black dots show more or less the same stars from the $N$-body simulation as shown in Fig.~\ref{1202_sim_circular}. Here stars within a cluster radius of 40 pc were again omitted for clarity. As in Fig.~\ref{1202_sim_circular} the blue solid lines show the streaklines for which the cluster
    mass was neglected, whereas for the red solid lines the cluster mass was taken into account. Also in velocity space the shape of the tails is much better reproduced by the streaklines which take the cluster mass into account.}
  \label{1202_vel_circular}
\end{figure}
The three $N$-body computations which we use here for comparison with
streaklines are taken from \citet{Kuepper10a}. The models were set-up using
the publicly available code
\textsc{McLuster}\footnote{\texttt{www.astro.uni-bonn.de/\~{}akuepper/mcluster/mcluster.html\\or
    www.astro.uni-bonn.de/\~{}webaiub/german/downloads.php}}
\citep{Kuepper11b}. They initially consist of 65536 stars drawn from the
canonical initial mass function  (\citealt{Kroupa01}, eq.~2) with masses between
$0.1\msun$ and $1.2\msun$, resulting in a cluster mass of about
$20000\msun$. The cluster stars follow a Plummer density profile with a
half-mass radius of 8 pc.

The clusters are initially located at their apocentre of 8.5 kpc within an
Milky Way-like potential (\citealt{Allen91}; see Sec.~\ref{sec:models}). We chose three orbital
eccentricities for the clusters of $\epsilon = \{0.00, 0.25, 0.50\}$, which
start with an orbital velocity of $v_{orb} = \{220, 165, 110\}$ kms$^{-1}$.
From Eq.~\ref{eq:rtc} we can infer that the tidal radius of these
  clusters is of the order of 40 pc. Hence, with a ratio of half-mass radius
  to tidal radius of about 0.2 the models belong to the group of extended,
  tidally filling clusters as categorised by \citet{Baumgardt10}.

The computations were performed with \textsc{Nbody4} \citep{Aarseth03} on the
\textsc{Grape} special-purpose computers \citep{Fukushige05} at AIfA Bonn for
4 Gyr. Here we use snapshots in the middle of these computations at different
orbital phases as shown in Fig.~\ref{orbit} and summarised in
Tab.~\ref{table1}.

\subsection{Circular orbit}
The first snapshot (Fig.~\ref{1202_sim_circular}) is from the cluster on a
circular orbit after 2 Gyr of dynamical evolution. The star cluster is shown
in projection onto the orbital plane similar to Fig.~\ref{streak_circular}. It
has lost about $3600\msun$, of which most stars escaped into the tidal tails
(black dots in Fig.~\ref{1202_sim_circular}; stars within a cluster radius of
40 pc are not shown for clarity). The predicted over- and underdensity pattern
is clearly visible in both the leading and the trailing tail. Moreover, we see
that the tidal tails have a rolling shape, resulting from the epicyclic motion
of the tail stars.

The tidal radius, $x_L$ (Eq.~\ref{eq:rt}), of the cluster at this moment is 37
pc, so from Eq.~\ref{eq:yc2} we expect the length of the epicycles, $y_C$, to
be about 350 pc when the stars escape with zero tangential velocity, $v_t$.
We find, however, that they are closer to 400-500 pc. In \citet{Kuepper10a} we
argued that this is due to the cluster stars escaping with small excess
velocities of about 0.27-0.55 kms$^{-1}$ in the direction of the tails. This
is a reasonable assumption, since such excess velocities have been measured
for evaporating stars in $N$-body simulations. \citet{Kuepper08b} found the
velocities of evaporating stars to be of a log-normal distribution with a peak
at roughly the mean stellar velocity within the cluster. Also, it is clear from Fig.~\ref{streak_circular} that, in our simple model, the distance between overdensities depends on the velocity with which escapers are
released.  Nevertheless the streaklines shown in Fig.~\ref{1202_sim_circular} were generated
from a single value of the initial velocity, which we have found to
produce a satisfactory fit to the structure of the tidal tails.

The streaklines of these assumed escape conditions are shown in blue in
Fig.~\ref{1202_sim_circular}. As we can see, the epicyclic pattern is well
reproduced, that is the epicycles have the correct length. Nevertheless, these
streaklines do not seem to perfectly describe the motion of escaping stars. It
seems like most escapers rather follow trajectories which are less curved and
which are closer to the cluster orbit on average. However, we neglected the
cluster mass for these streaklines.

For the red solid lines in Fig.~\ref{1202_sim_circular} the cluster mass was
taken into account. The test particles were released with zero velocity from
the tidal radius in this case. The streaklines seem to perfectly match the
mean path of the escaping stars,  therefore they reflect the dominant mode of escape. The scatter about this mean path may well be due to the scatter in escape conditions mentioned above.

In Fig.~\ref{1202_vel_circular} the comparison between the $N$-body data and our simple streakline model is shown as in Fig.~\ref{1202_sim_circular} but now in velocity space. The epicyclic motion of tail stars can also be seen as a periodic velocity variation. Like in Fig.~\ref{1202_sim_circular} we can see that the $N$-body data is reproduced better by the streaklines for which the cluster mass is taken into account. Our simple model therefore seems to represent the average escape conditions. As can be seen in Figs.~\ref{1202_sim_circular} \& \ref{1202_vel_circular}, the deviations from these average escape conditions are relatively small.

\subsection{Eccentric orbits}
\begin{figure*}
\includegraphics[width=168mm]{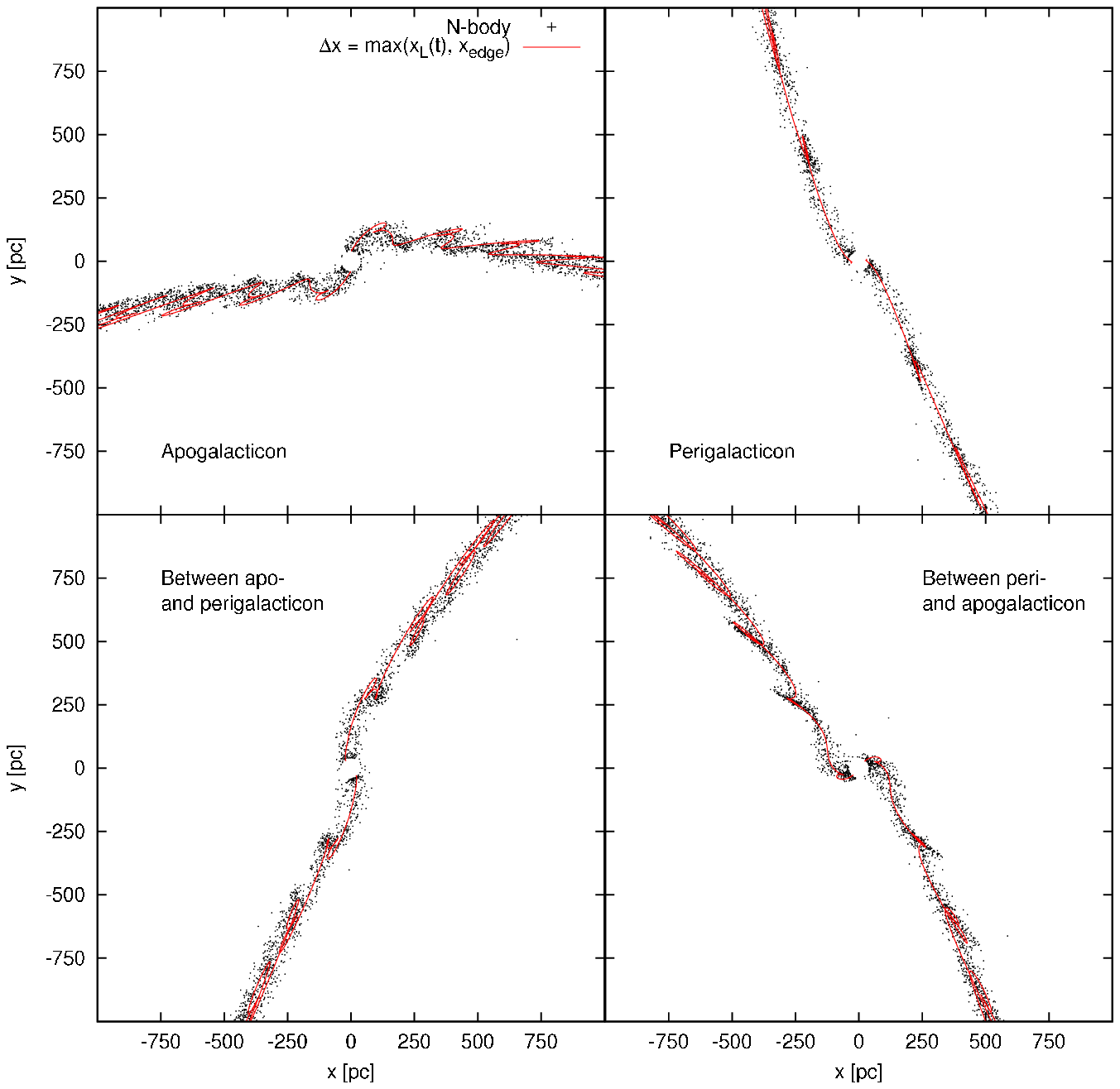}
  \caption{The same as Fig.~\ref{1202_sim_circular} but for the cluster with
    an orbital eccentricity of 0.25. This cluster has lost already $5000\msun$
    of stars (black dots; stars within a cluster radius of 40 pc were omitted
    for clarity). As in Fig.~\ref{1204_streak} \& \ref{1204_streak2}, the four
    snapshots show the cluster in four different orbital phases. The shape and
    density of the tidal tails changes significantly during one orbit. At
    apogalacticon they look similar to the circular orbit case but squeezed
    together. At perigalacticon the tails are stretched and the overdensities
    are much further away from the cluster than at apogalacticon. The orbital
    phases in between look quite distinct: between peri- and apogalacticon the
    tails show a streaky structure which has often been observed in numerical
    simulations. Also shown are streaklines for which the test particles have
    been released from the actual tidal radius (Eq.~\ref{eq:xoft}) with zero
    velocity (red solid lines). The edge radius which was used here is 32.5
    pc, in comparison to the  24.8 pc perigalactic tidal radius and the  39.5 pc
    apogalactic tidal radius. This set-up can reproduce all four snapshots
    without further ado, and especially reproduces the streaky features
    between peri- and apogalacticon. Such features have often been interpreted
    as results of tidal shocks in the past.}
  \label{1204_sim}
\end{figure*}
\begin{figure*}
\includegraphics[width=168mm]{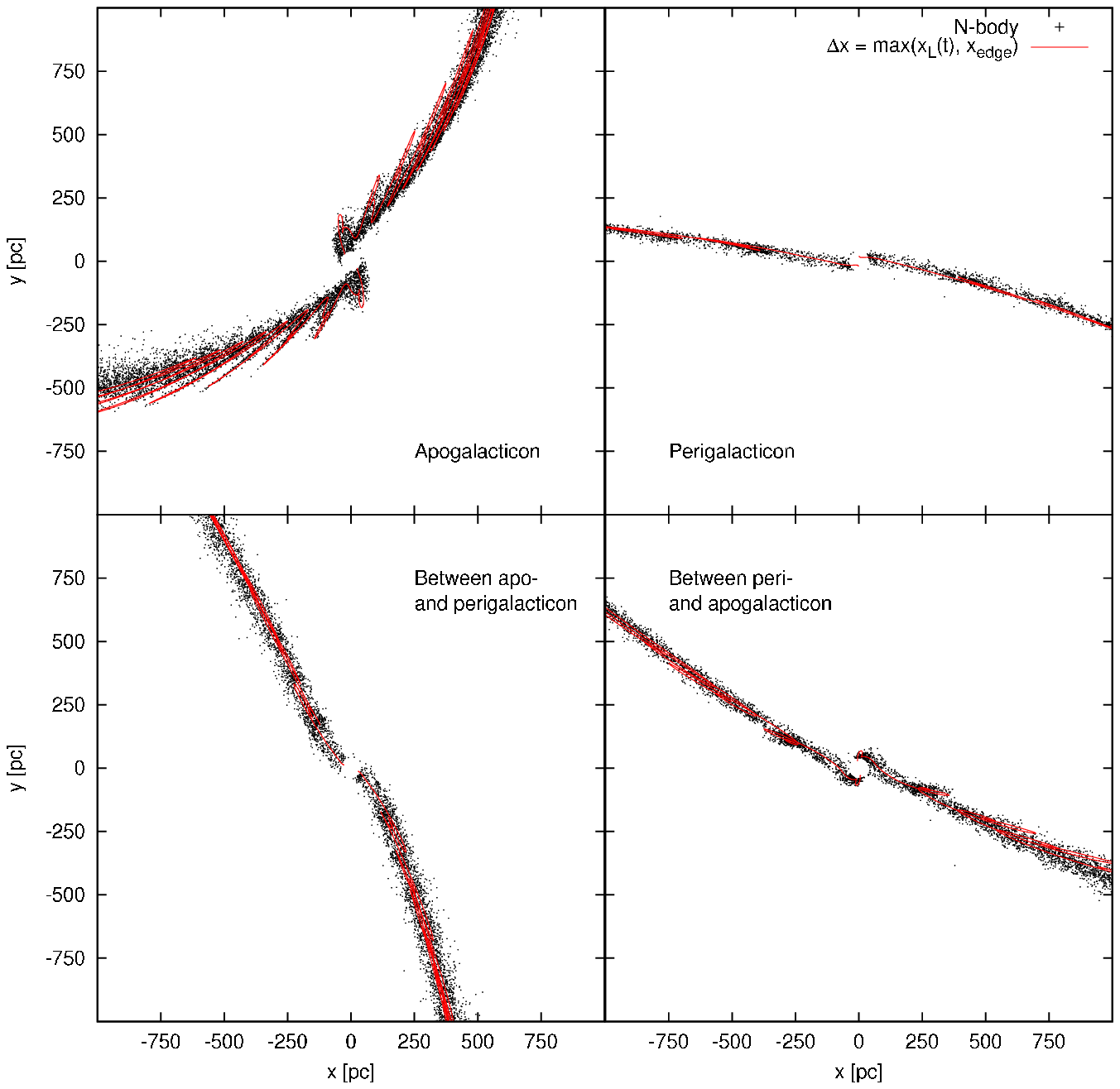}
  \caption{The same as Fig.~\ref{1202_sim_circular} \& \ref{1204_sim} but for
    an orbital eccentricity of 0.5. Due to the high eccentricity, this cluster
    has already lost $10500\msun$ of stars (black dots; stars within a cluster
    radius of 40 pc were omitted for clarity). Here, like in
    Fig.~\ref{1203_streak}, the four snapshots show four different orbital
    phases. We see that the compression and stretching of the tails gets more
    pronounced with increasing orbital eccentricity. Furthermore, at
    apogalacticon the compressed epicyclic loops appear like streaky
    features. These structures can be well reproduced by streaklines, when we
    release the the test particles from the actual tidal radius (red solid
    lines). The edge radius we used here is 18.0 pc, in comparison to the  12.9
    pc perigalactic tidal radius and the  36.4 pc apogalactic tidal radius.}
  \label{1203_sim}
\end{figure*}
In Fig.~\ref{1204_sim} the four consecutive snapshots of the cluster on an
orbit with an eccentricity of 0.25 are shown. The cluster has lost about
$5000\msun$ of stars which are mainly in the tidal tails (black dots). Here,
again, the epicyclic movement of the tail stars is visible. The tails show
periodic overdensities with a distance from the cluster which varies with the
orbital period (as observed by \citealt{Kuepper10a}). At apogalacticon the
overdensities are close to the cluster, whereas they are furthest away at
perigalacticon.

Also the shape of the tails depends on the orbital phase. At apogalacticon
they appear similar to the circular orbit case with the rolling
shape. However, the tails are denser and broader as they are being compressed
due to differential acceleration across the tails. At perigalacticon the tails
are thinnest and they appear almost like straight lines. The two snapshots in
between appear quite different even though the cluster is in a similar orbital
phase, only being accelerated in the one snapshot and decelerated in the
other. The distance of the overdensities is comparable, whereas the shape of
the tails is very distinct. In between peri- and apogalacticon the tails show
prominent streaks like have often been observed in numerical simulations
before  \citep{Dehnen04, Capuzzo05, Montuori07}, but which have not
  yet been assigned to epicyclic motion of tail stars. These features may also
  be the true nature of the multiple tidal tails observed for Galactic
  satellites like NGC 288, Willman 1 and NGC 2298 \citep{Leon00, Willman06,
    Balbinot11}.

Also shown are streaklines for which the test particles have been released
from the actual tidal radius (Eq.~\ref{eq:rt}). However, since the tidal
radius at perigalacticon is very small  (24.8 pc) and the passage through
perigalacticon is quite fast, there are many test particles which get
re-captured from the cluster as it moves to apogalacticon. Therefore, we had
to introduce a minimum radius, $x_{edge}$, from which test particles are released
(see Sec.~\ref{sec:streaklines},  Eq.~\ref{eq:xoft}). This ``edge'' radius was found to be 32.5 pc
by incrementing its size from zero until no test particles was
re-captured. Its size is about 80\% of the cluster's apogalactic tidal
radius (39.5 pc). Remarkably, with this minimal choice of edge radius the
streaklines reproduce the $N$-body data very well, even though their shape is
very sensitive to this value.

In Fig.~\ref{1203_sim} the four consecutive snapshots of the cluster on an
orbit with $\epsilon = 0.5$ are shown. This cluster has already lost
$10500\msun$ of stars which are now mainly in the tidal tails (black
dots). The orbital compression of the tails gets stronger at
apogalacticon. The overdensities are very close to each other and the tails
are very broad. Moreover, they show prominent streaky features. At
perigalacticon the tails become even thinner than in the $\epsilon = 0.25$
case. From the snapshot between peri- and apogalacticon we can see that the
cluster-tail system is experiencing severe differential acceleration, such
that the leading tail (to the right) is already being broadened, whereas the
trailing tail is still quite thin.

The red solid lines in Fig.~\ref{1203_sim} have been produced by releasing
test particles from the actual tidal radius and assuming a minimum tidal
radius (edge radius) of  18.0 pc  (Eq.~\ref{eq:xoft}). The perigalactic tidal radius of this cluster
is  12.9 pc and the apogalactic tidal radius is  36.4 pc.  Hence, the edge radius is about 50\% of the apogalactic tidal radius. Again, as for the
cluster with an orbital eccentricity of 0.25, this choice of escape conditions
reproduces the $N$-body data remarkably well.

Finally, from Fig.~\ref{1204_sim} \& \ref{1203_sim} it is obvious that all
substructure within the tidal tails can be ascribed to epicyclic motion of
tail stars. This rules out tidal shocks as origin of substructure in these
systems. Moreover, it demonstrates how complex the motion of stars within tidal tails of star clusters like Pal 5 can be, and thus how unlikely the formation of overdensities in such tidal tails through Jeans instabilities is \citep{Quillen10, Schneider11}.

\section{Summary and conclusions}\label{Sec:Conclusions}
We have investigated the motion of stars escaping from star clusters using
streaklines (for details on streaklines see Sec.~\ref{sec:streaklines}),
focussing on three star cluster models with orbital eccentricities of 0.0,
0.25 and 0.5 (further description of the models in Sec.~\ref{sec:models}). We
demonstrated how escaping stars move on epicycles within tidal tails
(Sec.~\ref{sec:circular} \& \ref{sec:eccentric}), and that this epicyclic
motion is the only origin of the substructure observed in $N$-body
computations of dissolving star clusters (Sec.~\ref{Sec:Numerical}).

In \citet{Kuepper08a} it was shown analytically and numerically for circular
cluster orbits, how the length of the epicycles changes with the size of the
radial offset, $\Delta x$, escaping stars have with respect to the cluster orbit at
the moment of escape. Here we visualised, with the help of streaklines, how
the shape of the epicycles changes when we vary the tangential velocity,
$v_t$, of the escaping stars. In fact, the shape of the epicycles and thus the tidal tails is very sensitive to this velocity. Furthermore, we demonstrated how this motion
gets distorted when the cluster orbit is eccentric, and what (important)
influence the cluster mass has.

When a cluster on an eccentric orbit is close to apogalacticon then
  the epicyclic motion of stars within its tails may lead to streaks appearing
  like multiple tails. Such features have been observed in numerical
  simulations before \citep{Dehnen04, Capuzzo05, Montuori07}, but have not yet
  been interpreted as result of epicyclic motion. The streaks may also be the
  true nature of the multiple tidal tails observed for Galactic satellites
  like NGC 288, Willman 1 and NGC 2298 \citep{Leon00, Willman06, Balbinot11}.

Finally, we compared sets of streaklines to results from three $N$-body models
following the same orbits as stated above. We found that the tidal tails of
these computations can be accurately reproduced by a quite simple model, that is when we assume that the stars evaporate from the actual tidal radius, $x_L(t)$, the star cluster has at the moment when the star escapes. When the velocity of the escapers is chosen such that they have the same angular velocity as the cluster, the shape of the tidal tails is reproduced best.  This is what we find to be the dominant mode of escape.

However, we have to assume that there is a lower limit, $x_{edge}$, to the tidal radius. That is, stars escape from $\Delta x(t) = \max (x_L(t), x_{edge}$). This ``edge'' radius was found to be identical to the tidal radius of the cluster, if it moves on a circular orbit, but was found to decrease with increasing orbital
eccentricity  at fixed apocentric distance. For all orbits, the edge radius turned out to be significantly
larger than the corresponding perigalactic tidal radius. It can be interpreted
as the minimum radius from which a star can escape from the cluster during one
cluster orbit without being re-captured by the cluster in a later orbital
phase. 

 These findings have only been established, however, for systems where the escape of stars is dominated by collisional heating, on galactic orbits with eccentricities up to 0.5.  For much larger eccentricities,
or any other system where mass loss may be dominated by tidal heating, modelling the structure of tidal tails may require refinement, as suggested by the collisionless modelling in \citet{Penarrubia09}. Even in the case of a cluster on a circular galactic orbit, the notion of a single escape radius is over-simplistic \citep{Read06}, but even these authors find that, in the long run, the ``tidal radii'' of stars on prograde, circular and retrograde orbits tend to a single value.

\citet{Kuepper10b} investigated in detail the same three $N$-body models that
we discuss in the present paper. They found that all three models show a
radial density profile which barely changes with time. By fitting King
profiles \citep{King62} and a new, so-called, KKBH profile, they showed that
the fitted tidal radius does not change with the orbital period, even though
Eq.~\ref{eq:rt} would suggest that the tidal radius should be large at
apogalacticon and small at perigalacticon. For this reason they suggested to
rather name such a fitted tidal radius the ``edge'' radius. Furthermore, they
found that the fitted edge radius of the model on a circular orbit is close to
its true tidal radius of 37 pc, whereas the model with an orbital eccentricity
of 0.25 showed an edge radius of about 30 pc, and the one with $\epsilon =
0.5$ had a constant edge radius of about 20 pc. These values lie remarkably
close to the values we found by adjusting streaklines to the tidal tails of
these $N$-body models.

These results suggest that the edge radius is fundamentally connected to the
orbit of a star cluster and its mass. Stars preferentially evaporate from the
actual tidal radius of the cluster whose lower limit is given by the edge
radius.  That is, stars within the edge radius cannot escape, whereas stars outside the edge radius get ``eaten away'' by the tide. This edge radius can be recovered by fitting a King or a KKBH profile
to the radial density profile. It must be mentioned, though, that
  these findings will most likely only hold for extended, tidally filling
  clusters (see \citealt{Baumgardt10}) with a ratio of half-mass radius to
  tidal radius larger than approximately 0.1, since more concentrated clusters
  may not have properly adapted to the mean tidal field yet. 
  
 Furthermore, applying our results to dwarf spheroidal galaxies may be complicated by the fact that in such systems the crossing time of stars orbiting at large radii may easily exceed the orbital time of the dwarf galaxy about its host. Instead, \citet{Penarrubia09} find that in dwarf galaxies which experienced a recent tidal  shock an edge develops which moves radially outward with time. Such a dwarf galaxy will lose most of its mass due to tidal stripping rather than through dynamical evaporation. The scatter in escape conditions of escaping stars would be much higher in this case so that our simple model would not work for the majority of escapers. Thus, no pronounced epicyclic signature would be expected for the tidal tails of such objects.

Originally, \citet{King62} suggested that the observed cut-off radius,
  i.e.~edge radius, of a star cluster on an eccentric cluster orbit is related
  to its respective perigalactic tidal radius
 \begin{equation}\label{eq:rtecc}
x^{peri}_{t} = \left(\frac{GM}{(2+\epsilon)\Omega_{peri}^2} \right)^{1/3},
 \end{equation}
where $\Omega_{peri}$ is the cluster's angular velocity at
perigalacticon. Hence, the cut-off radius should be even significantly smaller
than the perigalactic tidal radius for high eccentricities. Similar results
have been found by \citet{Read06} and \cite{Kennedy11}, both predicting the
cut-off radius to be of the order of the perigalactic tidal radius. In
contrast to this, we found the edge radius of clusters on eccentric orbits to
be significantly larger than the perigalactic radius. The reason for this
difference is that stars that become unbound at perigalacticon at smaller
radii get re-captured by the cluster at a later orbital phase.

Finally we conclude that, if sufficient information on a specific globular cluster of the Milky Way is available, this knowledge on the formation of tidal debris can be used to predict the shape of its tidal tails and the location of its epicyclic overdensities. In a future investigation we are going to demonstrate this with the globular cluster 47 Tucanae  (Lane, K\"upper \& Heggie, in prep.). Moreover, we would like to point out that epicyclic motion of stars within tidal tails of star clusters leads to over- and underdensities independent from the cluster orbit about the host galaxy. Therefore substructure in tidal tails is not necessarily due to substructure in the galactic potential like spiral arms, giant molecular clouds or dark-matter subhaloes (e.g.~\citealt{Ibata02, Carlberg09}).

\section*{Acknowledgments}
The authors are grateful to Sverre Aarseth for making his \textsc{Nbody} codes accessible to the public. Moreover they would like to thank Jorge Pe\~narrubia for useful comments. AHWK kindly acknowledges the support of an ESO Studentship and through the German Research Foundation (DFG) project  KR 1635/28-1. RRL acknowledges support from the Chilean Center for Astrophysics, FONDAP Nr.~15010003, the BASAL Centro de Astrof\'isica y Tecnolog\'ias Afines (CATA) PFB-06/2007 and the GEMINI-CONICYT Fund, allocated to project No.~32090010.

\bsp

\label{lastpage}
\end{document}